\documentclass[prb,twocolumn]{revtex4}
\usepackage{graphicx}
\usepackage{dcolumn}
\usepackage{bm}

\begin{document}

\title{Renormalization approach for quantum-dot structures under strong alternating
fields}
\author{P.~A.~Schulz}
\affiliation{Instituto de F\'{\i}sica `Gleb Wataghin', Universidade Estadual de Campinas,
13083--970, Campinas, S\~ao Paulo, Brazil}
\author{P.~H.~Rivera}
\affiliation{Departamento de F\'{\i}sica, Universidade Federal de S\~ao Carlos,
13565--905, S\~ao Carlos, S\~ao Paulo, Brazil}
\author{Nelson~Studart}
\affiliation{Departamento de F\'{\i}sica, Universidade Federal de S\~ao Carlos,
13565--905, S\~ao Carlos, S\~ao Paulo, Brazil}

\begin{abstract}
We develop a renormalization method for calculating the electronic structure
of single and double quantum dots under intense ac fields. The
nanostructures are emulated by lattice models with a clear continuum limit
of the effective-mass and single-particle approximations. The coupling to
the ac field is treated  non-perturbatively by means of the Floquet
Hamiltonian. The  renormalization approach allows the study of dressed
states of  the nanoscopic system with realistic geometries as well arbitrary
strong ac fields. We give examples of a single quantum dot, emphasizing the
analysis of the effective-mass limit for lattice models, and double-dot
structures, where we discuss the limit of the well used two-level
approximation.
\end{abstract}


\maketitle

\section{Introduction}

The renormalization method, as a tool for studying the electronic structure
of solids, has attracted attention already twenty years ago with the
application to disordered one-dimensional systems.\cite{aoki80,dasilva81} It
is a powerful approach for systems whose complexity hinders the direct
diagonalization of the corresponding Hamiltonian. At that time its complete
usefulness could be hardly realized due to the difficulty in applying the
method simultaneously in more than one spatial direction. More recently its
suitability has been revealed for studying strongly anisotropic solids that
started to deserve growing interest, like semiconductor superlattices and
conducting polymers.\cite{farchioni99}

In the present work the renormalization method is applied to a different
problem of increasing interest with the same formal structure: the dressed
electronic spectra of quantum dot systems under intense ac fields.

In semiconductor technology, the continuous improvement on growth techniques
opened possibilities to design systems that are in the quantum limit in all
spatial directions, establishing an important and pioneering branch of what
is nowadays called nanoscience. One of these devices, a double-quantum dot
is a candidate for the elementary unit of the foreseen quantum computing.
Two coupled quantum dots afterwards called as \textit{artificial molecules}%
\cite{leo95} has been the object of intensive research both
theoretical,\cite{bryant91,klimeck94,niu95,palacios95,oh96,kotlyar97a,aono97,aoki97,poh97,brune97,mayrock97,kapu98,imamura98,asano98,xie98,stafford98,partoens01}
and
experimentally,\cite{lorke90,tewordt94,waugh95,dixon96,blick96,schmidt97,blick98a,blick98b,ooster98a,ooster98b,fuji98}
in the last few years.

Mesoscopic systems, like quantum dots and quantum dot arrays, are
described by lattice models, treated in a tight-binding framework,
emulating the continuum limit of the effective-mass
approximation.\cite{huckestein95} Lattice models have been used
mainly in the context of disorder effects on electronic and
transport properties of two-dimensional systems and quantum
billiards.\cite{louis01} On the other hand, there are reports in
the literature on using lattice models in another situation, where
they constitute also an appropriate approach, namely the
simulation of arrays of quantum dots and antidots in the presence
of a magnetic field.\cite{zozoulenko95} Realistic description of
quantum dot nanostructures in the independent particle
approximation interacting with intense ac fields
becomes possible within the proposed renormalization-decimation procedure.%
\cite{dasilva81}

The coupling to an ac field is non-perturbatively included using the Floquet
method, by means of a procedure introduced by Shirley.\cite{shirley} The
Floquet method has been widely used for the non-perturbative study of the
interaction of atomic, molecular and semiconductor nanostructures systems
with a strong ac electric field. The time-independent infinite matrix
Hamiltonian obtained after the application of the Floquet-Fourier
transformation over the time-dependent Schr\"odinger equation, describes
entirely these processes without any further \textit{ad-hoc} assumptions.

Therefore, the effect of an intense ac field on the electronic
spectra of a nanostructure, like a quantum dot, is well described
by an infinite Floquet matrix. However, due to the large vector
basis used to define the system in a lattice-model tight-binding
approximation, the necessary convergence criteria make the
eigenvalue calculations practically impossible beyond the
perturbative field intensity range. Moreover, we observe that the
infinite matrix Hamiltonian structure resembles the structure of a
linear chain matrix, where the ``energy sites'' are energy
sub-matrices which corresponds to the Hamiltonian of the system
plus or minus an associated multiple of the photon energy and the
``hopping parameter'' is the ac field coupling the sub-matrices.
These relationships allow us to develop an interesting and
promising renormalization approach whereby the actual dimension of
the system to be calculated is reduced to the one of the lattice
model for the bare system. Besides that, since the quantity
calculated is the density of states, one gets a step further than
by the direct diagonalization of the Floquet Hamiltonian, which
provides only the quasi-energy spectra without the spectral
modulation as a function of the field strength. The hierarchy of
these quasi-energy spectra related to different photon replicas
will also be discussed.

The paper is presented as follows. A brief introduction to the
Floquet Hamiltonian is given in Sec. II, followed by a
comprehensive description of the renormalization method applied to
the system. In Sec. III, numerical calculations for single and
double quantum dots are shown and discussed. We first focus on the
validity of a lattice model for a quantum dot in the presence of
intense ac fields. The second part is centered on the ac field
dependence of the bonding and antibonding states of an artificial
molecule, i.e., a double-quantum dot structure. In Sec. IV we
present our conclusions.

\section{Theory}

\subsection{The Floquet Hamiltonian}

The lower part of the energy spectrum of a mesoscopic system, like a quantum
dot or quantum-dot array described in the framework of the effective-mass
approximation, will be evaluated here using a tight-binding model for a
square lattice of \textit{s}-like orbitals, considering only
nearest-neighbors interaction and a hopping parameter defined by $%
V=-\hbar^2/(2m^*a^2)$, where $m^*$ is the effective mass and $a$ is the host
lattice parameter.

The applied ac fields are parallel to one of the square sides. Hence, our
model is described by the Hamiltonian $H= H_0+H_{int}$ where

\begin{displaymath}
H_o=\sum_{l_1,l_2}\epsilon_{l_1,l_2}\mathbf{\sigma}_{l_1,l_2}\mathbf{\sigma}_{l_1,l_2}^{\dag}+{V\over2}\sum_{l_1,l_2}\Bigl[\mathbf{\sigma}_{l_1,l_2}\mathbf{\sigma}_{l_1+1,l_2}^{\dag}+
\end{displaymath}
\begin{equation}
\mathbf{\sigma}_{l_1+1,l_2}\mathbf{\sigma}_{l_1,l_2}^{\dag}+\mathbf{\sigma}_{l_1,l_2}\mathbf{\sigma}_{l_1,l_2+1}^{\dag}+\mathbf{\sigma}_{l_1,l_2+1}\mathbf{\sigma}_{l_1,l_2}^{\dag}\Bigr]
\end{equation}

\noindent and
\begin{equation}
H_{int}=eaF\cos \omega t\sum _{l_1,l_2}\mathbf{\sigma}_{l_1,l_2}l_1\mathbf{%
\sigma}_{l_1,l_2}^{\dag}
\end{equation}

\noindent with $\mathbf{\sigma}_{l_1,l_2}=|l_1,l_2>$ and $\mathbf{\sigma}%
_{l_1,l_2}^{\dag}=<l_1,l_2|$. The \textit{atomic energies} will be taken
constant, $\epsilon_{l_1,l_2}=4|V|$, for all sites. $F$ and $\omega$ are the
ac field amplitude and frequency respectively, and $e$ is the electron
charge. The treatment of the time-dependent problem is based on Floquet
states $|l_1,l_2,m>$ where $l_1,l_2$ are the site indexes and $m$ is the
photon index. We follow the procedure developed by Shirley\cite{shirley}
which consists in a Fourier-Floquet transformation of the time-dependent
Hamiltonian into a time-independent infinite matrix which must be truncated.
The matrix elements are

\begin{displaymath}
\Bigl[(\mathcal{E}-m\hbar\omega
-\epsilon_{l_1,l_2})\delta_{l_1^{\prime}l_1}\delta_{l_2^{\prime}l_2}-{V\over2}\Bigl\{(\delta_{l_1^{\prime},l_1-1}+\delta_{l_1^{\prime},l_1+1})\delta_{l_2^{\prime}l_2}\Bigr.\Bigr.
\end{displaymath}
\begin{displaymath}
\Bigl.\Bigl.+(\delta_{l_2^{\prime},l_2-1}+\delta_{l_2^{\prime},l_2+1})\delta_{l_1^{\prime}l_1}\Bigr\}\Bigr]\delta
_{m'm}
\end{displaymath}
\begin{equation}
=F_1l_1\delta_{l_1^{\prime}l_1}\delta_{l_2^{\prime}l_2}(\delta
_{m^{\prime},m-1}+\delta _{m^{\prime},m+1})
\end{equation}

\noindent where $F_1={\frac{1}{2}}eaF$. The dimension of the matrix is $%
L_1\times L_2(2M+1)$, where $L_1$ and $L_2$ are the maximum number of atomic
sites, while $M$ is the maximum photon index. We choose $M$ in order to
satisfy a convergence condition: symmetric spectra relative to the edges of
the quasi Brillouin zones (QBZs). The first QBZ is spanned in the range $%
-\hbar\omega/2\leq \mathcal{E}\leq\hbar\omega/2$\ .

The truncated Floquet matrix resembles a \textit{linear chain} matrix in the
form

\begin{equation}
\label{eq:floquet1} \left(
\begin{array}{*{10}{c}}
\mathsf{E}^M & \mathcal{F} \cr \mathcal{F} & \mathsf{E}^{M-1} &
\mathcal{F} \cr & & \ddots \cr & & \mathcal{F} & \mathsf{E}^{1}
& \mathcal{F} \cr & & & \mathcal{F} & \mathsf{E}^{0} & {\mathcal
F} \cr & & & & \mathcal{F} & \mathsf{E}^{-1} & \mathcal{F} \cr &
& & & & \ddots \cr & & & & & & \mathcal{F} & \mathsf{E}^{-M+1} &
\mathcal{F} \cr & & & & & & & \mathcal{F} & \mathsf{E}^{-M} \cr
\end{array}
\right)
\end{equation}

\noindent where $\mathsf{E}^m={\mathcal
E}-(\mathsf{S}+m\hbar\omega)$ is a $L_1\times L_2$ matrix, with
$\mathcal{E}$ being the quasi-energy spectrum and $\mathsf{S}$,
is the $L_1\times L_2$ system matrix. For other hand, the
$L_1\times L_2$ diagonal matrix $\mathcal{F}$ represents the
coupling of the system with the ac electric field and is given by

\begin{equation}
{\mathcal
F}=F_1l_1\delta_{l_1^{\prime}l_1}\delta_{l_2^{\prime}l_2} \  \  .
\label{eq:field1}
\end{equation}

A time-independent Schr\"odinger equation defined in terms of Green
functions as\cite{economou83}

\begin{equation}  \label{eq:green}
(\mathcal{E}-H)G=1
\end{equation}
can be associated to the Floquet matrix, as we will show next.

\subsection{Renormalization Method}

Equation (\ref{eq:green}) can be projected to the photon Hilbert spaces $<n|$
and $|m>$, resulting in

\begin{equation}  \label{eq:renor1}
\sum_k<n|(\mathcal{E}-H)|k>G_{km}=\delta _{nm} \ .
\end{equation}

\noindent Expanding this equation, considering $\mathcal{E}\rightarrow
\mathcal{E}+i\eta$, where $\eta\rightarrow 0$, eliminating the intermediate
index \textit{sites}, making some substitutions, and repeating the
decimation process successively we arrive to the general expression, for any
$\xi\ge 1$ order decimation of the Eq.~(\ref{eq:renor1}):

\begin{equation}  \label{eq:renor2}
E_n^{\xi}G_{n0}=\delta_{n0}+\mathcal{F}_{n-2^{\xi},n}^{\xi}G_{n-2^{\xi},0}+\mathcal{F}%
_{n+2^{\xi},n}^{\xi}G_{n+2^{\xi},0}
\end{equation}

\noindent where

\begin{displaymath}
    E_n^{\xi}=E_n^{\xi-1}-\mathcal{F}_{n-2^{\xi-1},n}^{\xi-1}\frac{1}{E_{n-2^{\xi-1}}^{\xi-1}}\mathcal{F}_{n,n-2^{\xi-1}}^{\xi-1}-
\end{displaymath}
\begin{displaymath}
    \mathcal{F}_{n+2^{\xi-1},n}^{\xi-1}\frac{1}{E_{n+2^{\xi-1}}^{\xi-1}}\mathcal{F}_{n,n+2^{\xi-1}}^{\xi-1}
\end{displaymath}

\begin{displaymath}
\mathcal{F}_{n-2^{\xi},n}^{\xi}={\mathcal
F}_{n-2^{\xi-1},n}^{\xi-1}\frac{1}{E_{n-2^{\xi-1}}^{\xi-1}}{\mathcal
F}_{n-2^{\xi},n-2^{\xi-1}}^{\xi-1}
\end{displaymath}

\begin{displaymath}
\mathcal{F}_{n+2^{\xi},n}^{\xi}={\mathcal
F}_{n+2^{\xi-1},n}^{\xi-1}\frac{1}{E_{n+2^{\xi-1}}^{\xi-1}}{\mathcal
F}_{n+2^{\xi},n+2^{\xi-1}}^{\xi-1}  \  \  .
\end{displaymath}

The Green function $G_{00}$ after the first decimation is given by

\begin{equation}
  \label{eq:renor4}
\begin{array}{*{1}{r}}
G_{00}=\left[\left(E_{0}-\mathcal{F}\frac{1}{E_{-1}}{\mathcal
F}-\mathcal{F}\frac{1}{E_{1}}\mathcal{F}\right)-\right.\cr
\left.\left(\mathcal{F}\frac{1}{E_{-1}}{\mathcal
F}\right)\frac{1}{A_{-2}}\left({\mathcal
F}\frac{1}{E_{-1}}\mathcal{F}\right)-\right. \cr
\left.\left(\mathcal{F}\frac{1}{E_1}{\mathcal
F}\right)\frac{1}{A_2}\left(\mathcal{F}\frac{1}{E_1}{\mathcal
F}\right)\right]^{-1} \cr
\end{array}
\end{equation}

\noindent where
\begin{eqnarray*}
A_{-2}=\left(E_{-2}-\mathcal{F}\frac{1}{E_{-1}}\mathcal{F}-\mathcal{F}\frac{1}{E_{-3}}\mathcal{F}\right)- \\
\left(\mathcal{F}\frac{1}{E_{-3}}{\mathcal
F}\right)\frac{1}{B_{-4}}\left({\mathcal
F}\frac{1}{E_{-3}}\mathcal{F}\right);
\end{eqnarray*}
\begin{eqnarray*}
A_{2}=\left(E_{2}-\mathcal{F}\frac{1}{E_{1}}\mathcal{F}-\mathcal{F}\frac{1}{E_{3}}\mathcal{F}\right)- \\
\left(\mathcal{F}\frac{1}{E_{3}}\mathcal{F}\right)\frac{1}{B_{4}}\left(\mathcal{F}\frac{1}{E_{3}}\mathcal{F}\right) \\
\end{eqnarray*}

The density of states of Floquet is expressed as
\begin{equation}  \label{eq:dos}
\rho(\mathcal{E}+i\eta)=\lim_{\eta\rightarrow 0}\Big( -\frac{1}{\pi}~\mathrm{%
Im}\left[~\mathrm{Tr}~G_{00}~\right]\Big).
\end{equation}

\section{Artificial Atoms and Molecules}

\subsection{Quantum dot under intense ac field}

As a starting point we calculate the quasi-density of states of quantum dots
as a function of the ac field intensity. We choose a square and a rounded
geometry to represent the quantum dot as shown in Fig.~\ref{fig:atoms}. The
square quantum dot consists of an array of $6\times 6$ atomic sites, Fig.~%
\ref{fig:atoms}(a). A more realistic geometry is given in Fig.\ref{fig:atoms}%
(b), which will be used for the study of double-dot systems in
this work. It should be noticed that the area of this geometry (37
sites) is comparable with the square one if we use the same
tight-binding parameters given below. Such lattice models exhibit
a particle-hole symmetry in the electronic structure and are
usually thought as simple, although useful, approximations for
superlattices or arrays of quantum dots, where each quantum well
or quantum dot is represented by a lattice site, respectively.
This extreme lattice limit has been used for studying
qualitatively the effect of intense ac fields on superlattice
minibands.\cite{rivera00} On the other hand, lattice models may be
useful in calculating the lower part of electronic systems well
described by the effective mass approximation. In the present
work, the tight-binding hopping parameter is chosen in order to
emulate the
electron effective mass for the GaAs bottom of the conduction band, $%
m^*=0.067m_0$. Since $V=-\hbar^2/(2m^*a^2)$, $V=-0.142$ eV for a
model lattice parameter of $a=20$ \AA. This leads to a quantum dot
with a lateral width of $L=120$ \AA, an order of magnitude lower
than typical dimensions of actual quantum dots fabricated by
lithographic methods. We intend to illustrate the method with the
present calculations. Nevertheless, all conclusions can be traced
back to larger systems by an appropriate scaling of the energy
parameters.

The first point to be addressed here is establishing the limit between the
extreme-lattice and effective-mass limits. This is achieved by following the
evolution of the quasi-density of states as a function of the ac field
intensity with frequencies of the order of the electronic band width. In Fig.%
\ref{fig:dots} we show the contour plot of the density of states
as a function of the field intensity for the bottom half of the
first QBZ. We recall that, for the chosen hopping parameter, the
band width for a square array of single atomic $s$-like showing
particle-hole symmetry is given by $\Delta E = 8|V|= 1.136$ eV.
Here the field frequency is $\hbar\omega = 1.0$ eV. The square
quantum dot, even though a textbook example, is very useful to
understand the breakdown of the effective-mass limit. At zero
field intensity the energy spectrum of a 2D square potential well
with $L=120$ \AA\ - a quantum dot - is clearly identified, as well
as the breakdown of degeneracies due to the applied ac field. The
ac Stark shifts lead to crossings at $eaF/\hbar\omega\approx 2.4$,
corresponding to a dynamic localization along the chains
perpendicular to the field direction. This is the limit where the
host lattice effects are already  predominant, i.e., the discrete
basis of atomic orbitals does not emulate the  effective-mass
approximation anymore and the length scale is given by the  host
lattice parameter $a$ and not by the lateral width $L$ of the
quantum dot. Indeed, the model simulates in this limit a square
array of quantum dots,  each one represented by a single site,
resembling the spectra that would be expected for coupled chains,
where each chain mimics the dynamic localization in superlattice
minibands.

This result for a square array is a good guide for  characterizing the
density of states as a function of the field intensity of a  quantum dot
with lower symmetry, Fig.\ref{fig:atoms}(b), as can be seen in the
equivalent plot shown in Fig.\ref{fig:dots}(b), also for $\hbar\omega = 1.0$
eV.  The dynamic localization at the extreme lattice limit is  less defined
due to the fact that the atomic site chains perpendicular to the  field
direction are not equivalent as in the square array.  Furthermore some of
the degeneracies are already broken in the absence of the ac field.

One clear advantage of the method can already be seen in Fig.~\ref{fig:dots}%
. By diagonalizing the  Floquet Hamiltonian,  the quasi-energy
spectra are depicted in the so called QBZs, each of them
reproducing  the spectrum with the energy shifted by  integer
multiples of the photon energy. The overlap of these photon
replicas  makes the interpretation of the spectrum rather
cumbersome, specially for  strong overlaps, which are unavoidable
for low frequencies. The present renormalization  approach
indicates that the quasi-energy spectrum, modulated  by the field
dependent density of states for higher or lower photon replicas
becomes relevant only with
increasing field intensity. This effect is verified at the bottom of Figs.~%
\ref{fig:dots}(a) and \ref{fig:dots}(b). The spectrum replica lowered by one
photon energy shows a negligible contribution at low  field. On the other
hand,  in both figures we see no significant modulation of the density of
states in  the low field limit within the  entire band of the depicted zero
photon replica,  due to the high frequency considered.

The main interest, however, is the effective-mass limit, i.e., the energy
bottom of each photon replica at low fields in the scale of Fig.~\ref%
{fig:dots},  as well as low frequencies, that would couple only these few
low-energy states.  The scaling of these quantities shows the suitability of
the present method,  as will be seen in the following discussion on double
quantum dots.

\subsection{Double Quantum dot}

Our double-quantum dot system is based on the quantum dot shown in Fig.~\ref%
{fig:atoms}(b). The coupling between dots is of free choice and we
consider a simple connection of both dots by the same hopping
parameters considered so far, as illustrated in
Fig.~\ref{fig:mol1}. Such configuration is an example of strong
inter-dot coupling. The corresponding quasi-density of states plot
as a function of the field intensity for the same high field
frequency, $\hbar\omega = 1.0$ eV, as in Fig.~\ref{fig:dots} is
shown in Fig.~\ref{fig:mol2}(a). The quasi-density of states
spectrum is very similar with the appearance of the expected
energy splitting due to the coupling among quantum dot states. The
structure of the energy splitting may be rather complex,
considering the coupling of initially degenerate states in each
quantum dot. A covalent-like binding, with a splitting between a
bonding and antibonding states, is well defined for the inter-dot
coupling between the lowest state in each quantum dot. Therefore,
from now on we will focus exclusively on the energy bottom of the
spectra, the continuum low-field intensity limit, in order to
analyze the lowest pair of split double-quantum dot states.

In Fig.~\ref{fig:mol2}(b) we have the quasi-density of states for
the lowest pair of molecular states. Both bonding and antibonding
states shift rigidly upwards in energy. Here a modulation of the
density of states is seen. By increasing the field intensity, the
density of states diminishes, with increasing contribution of
higher and lower photon replica (not shown). Since the frequency
of the field is very high $\hbar\omega=1$ eV, all states of the
system are mixed by the field and no typical two-level behavior is
observed. The rigid energy shift may be seen already as a lattice
effect, since for high field intensities, the dynamic localization
for the host lattice is observed, Fig.\ref{fig:mol2}(a).

A clear covalent picture is revealed for much lower frequencies,
in the range of $\hbar\omega \approx 10$ meV, which is of the
order of the tunnel splitting between the lowest pair of states in
the bare ``molecule" for the chosen parameters:
$\Delta_{\text{split}}=7.1$ meV. For this frequency range the
coupling to higher molecular states is negligible, since the third
molecular state is about 50 meV above the antibonding state.
Lattice effects are also absent since the dynamic localization
effects on the host lattice are relevant for field frequencies of
the order of the entire spectral width, which is two orders of
magnitude larger than the energy scale of interest given by the
tunnel splitting. The present calculations are exact, since we are
in an independent particle approximation, for an artificial
$H_2^+$ molecule. Within this parameter range, the lowest pair of
states of the double-quantum dot behave as a two-level system, as
will be discussed next. We are going to switch the field frequency
from above to below the tunnel splitting energy. The quasi-density
of states plots will also be compared to the corresponding
quasi-energy spectra of the equivalent two level system, obtained
by diagonalizing the Floquet Hamiltonian.

In Fig. 5 the evolution of the artificial $H_2^+$ bonding and
antibonding states as a function of the field intensity is shown
for a frequency $\hbar\omega = 10 \ {\text{meV}} \  >
\Delta_{\text{split}}$.
 In Fig. 5(a) we see the quasi-density of
states around the main ``photon replica", which shows the highest
intensity for low fields. The next important branches of the
quasi-density of states are the bonding state plus one photon and
the antibonding minus one photon. With increasing the intensity,
the splitting between the zero-photon bonding and antibonding
states diminishes down to a crossing at $eaF/\hbar\omega \approx
0.3$. This resembles the dynamic localization regime for
superlattice minibands. \cite{rivera00} The branches of the
quasi-density of states in Fig. 5(a) can be mapped on the
quasi-energy spectrum of a two-level system, emulated by two
atomic sites with fitted tight-binding
parameters. This spectrum is a function of a field intensity defined by $%
edF/\hbar\omega$, where $d$ is the distance between the two
effective atomic sites. The correspondence between Figs. 5(a) and
5(b) is satisfied by properly scaling $d\approx7a$. Having in mind
the double quantum dot of Fig. 3, $d_0=7a$ is the distance between
the centers of the dots, i.e. ``the bond length of the molecule".

The equivalent situation, for a frequency given by $\hbar\omega =
7$ meV, therefore near the resonance situation
$\Delta_{\text{split}}\approx\hbar\omega$, is shown in Fig.6. Here
we clearly see that the bonding and antibonding states evolve in
\textit{Rabi sidebands}.\cite{haug90} This behavior can also be
mapped on an effective two-level system, Fig. 6(b). The effective
distance between the two effective atomic sites is close to the
``molecular bond length" $d_0=7a$, like for frequencies higher
than the tunnel splitting, Fig.5.

The quasi-resonant case may be analyzed directly from the Rabi
frequency at the crossing of the sidebands in Fig. 6(a). At the
crossing $\hbar\omega_R= \Delta_{\text{split}}$,
 where
$\omega_R=d_RF/\hbar$ with $d_R$ being the dipole matrix element.
Since
$\Delta_{\text{split}}=\hbar\omega$,
one has
that $eaF\approx0.3\hbar\omega_R$, resulting in a dipole matrix element $%
d_R\approx 3.33a$ and $2d_R=6.6a$, in good agreement with the
``molecular length" $d_0=7a$.

In Fig. 7 we show the situation for a field frequency $\hbar\omega
= 5$ meV lower than the bare tunnelling splitting. The zero-photon
molecular states show an ac Stark shift, Fig. 7(a). The dressed
tunnel splitting increases
with field intensity up to the first important anticrossing at $%
eaF/\hbar\omega =1$. Now the mapping on an effective two-level system,
Fig.7(b), occurs for an effective distance $d\approx3.5$, nearly half the
nominal molecular bond length. This shrinking of the effective distance is,
however, compatible with the increasing of the tunnel splitting.

\section{Conclusions}

The present results demonstrate the usefulness of the
renormalization-decimation method applied to the Floquet
Hamiltonian. Such a procedure makes possible the calculation of
quasi-density of states of realistic geometries for nanostructures
under intense ac fields described by lattice models. The matrix
dimension in the renormalization method is $L_1\times L_2$, while
the direct diagonalization has to handle with matrix dimension of
$(2M+1)\times (L_1\times L_2)$. This is of paramount importance if
we must consider $M\approx 100$ for $L_{1,2} > 20$ in order to
simulate realistic mesoscopic systems.\cite{rivera00} The examples
shown here for double quantum dots, although heuristic, point out
interesting dependence of the dressed electronic structure on the
field frequency. A natural extension of the work is the analysis
of the local quasi-density of states, as well as the study of
asymmetric double-quantum dots, a situation for which the density
of states reveals actual tunnelling  properties, in order to
discuss transport measurements in such systems.

\begin{acknowledgments}
It is a pleasure to thank H. S. Brandi for suggesting the
possibility of decimating the Floquet Hamiltonian, as well as many
encouraging discussions at the beginning of this work. The work
was supported by Funda\c{c}\~{a}o de Amparo \`{a} Pesquisa do
Estado de S\~{a}o Paulo (FAPESP) and Conselho Nacional de
Desenvolvimento Cient\'{\i}fico e Tecnol\'{o}gico (CNPq).
\end{acknowledgments}

\begin{figure}[htbp]
\begin{center}
\end{center}
\caption{Quantum dot geometries:(a) square and (b) rounded square
dots, where $a$ is the host lattice parameter.} \label{fig:atoms}
\end{figure}

\begin{figure}[htbp]
\begin{center}
\end{center}
\caption{Density of states for the (a) square and (b) rounded dots
as a function of the field intensity. The field frequency is
$\hbar\omega=1.0$ eV. Only the bottom half of the first QZB is
shown (see text).} \label{fig:dots}
\end{figure}

\begin{figure}[htbp]
\begin{center}
\end{center}
\caption{Double quantum-dot geometry based on rounded square dots,
where $d_0$ is the ``bond length".} \label{fig:mol1}
\end{figure}

\begin{figure}[htbp]
\begin{center}
\end{center}
\caption{Density of states of the double quantum-dot shown in Fig.\ref%
{fig:mol1} as a function of field intensity at high frequency
$\hbar\omega=1$ eV, (a) the bottom half of the first QBZ and (b)
the bonding and antibonding states.} \label{fig:mol2}
\end{figure}

\begin{figure}[htbp]
\begin{center}
\end{center}
\caption{(a) Density of states of a double quantum-dot as a function of the field intensity for a frequency, $%
\hbar\omega=10$ meV; (b) spectrum of a equivalent two-level
system, simulated with a dipole distance $d\approx d_0$, in the
dynamic localization regime.} \label{fig:mol3}
\end{figure}

\begin{figure}[htbp]
\begin{center}
\end{center}
\caption{(a) Same as Fig.~\ref{fig:mol3} for $\hbar\omega=7$ meV,
and (b) same as Fig.~\ref{fig:mol3} near the Rabi resonance.}
\label{fig:mol4}
\end{figure}

\begin{figure}[htbp]
\begin{center}
\end{center}
\caption{(a) same as Fig.~\ref{fig:mol3}for $\hbar\omega=5$ meV,
and (b) same as Fig.~\ref{fig:mol3} in the ac Stark regime.}
\label{fig:mol5}
\end{figure}

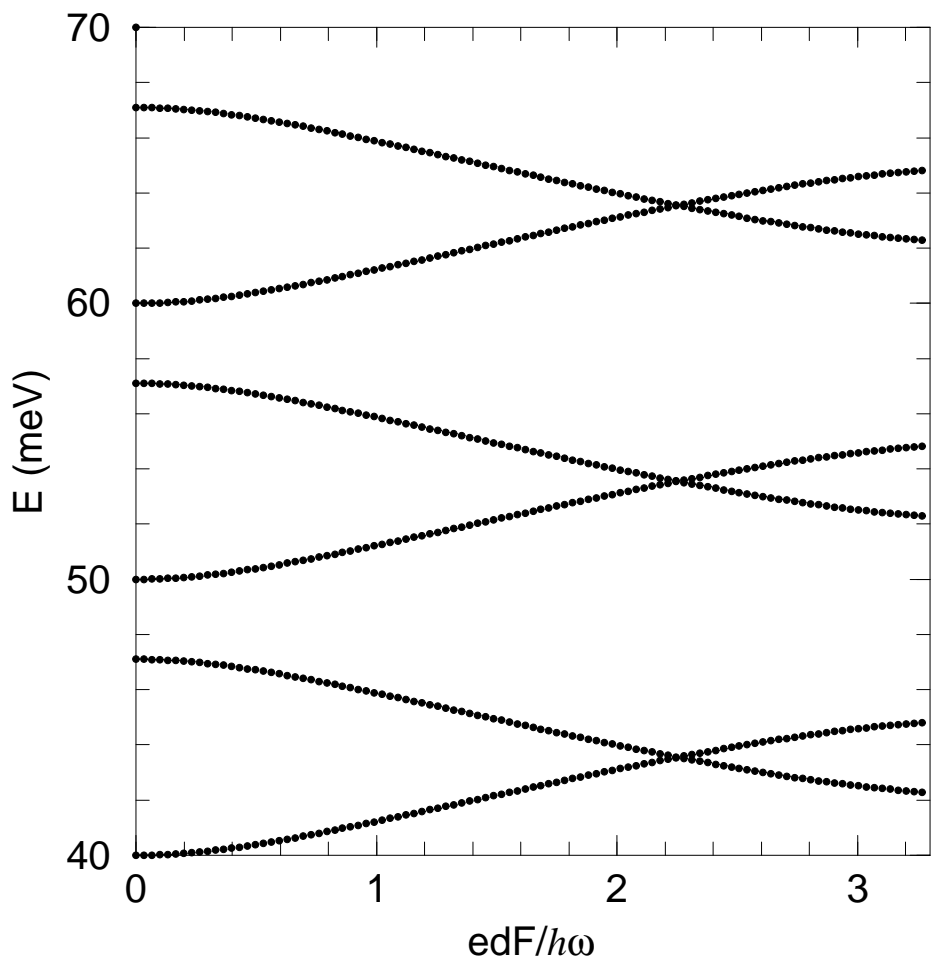
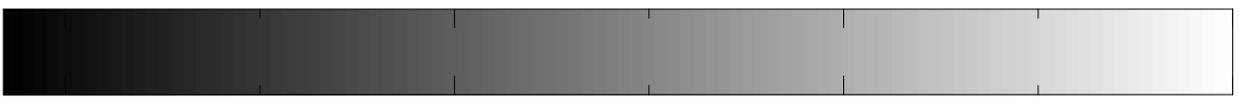
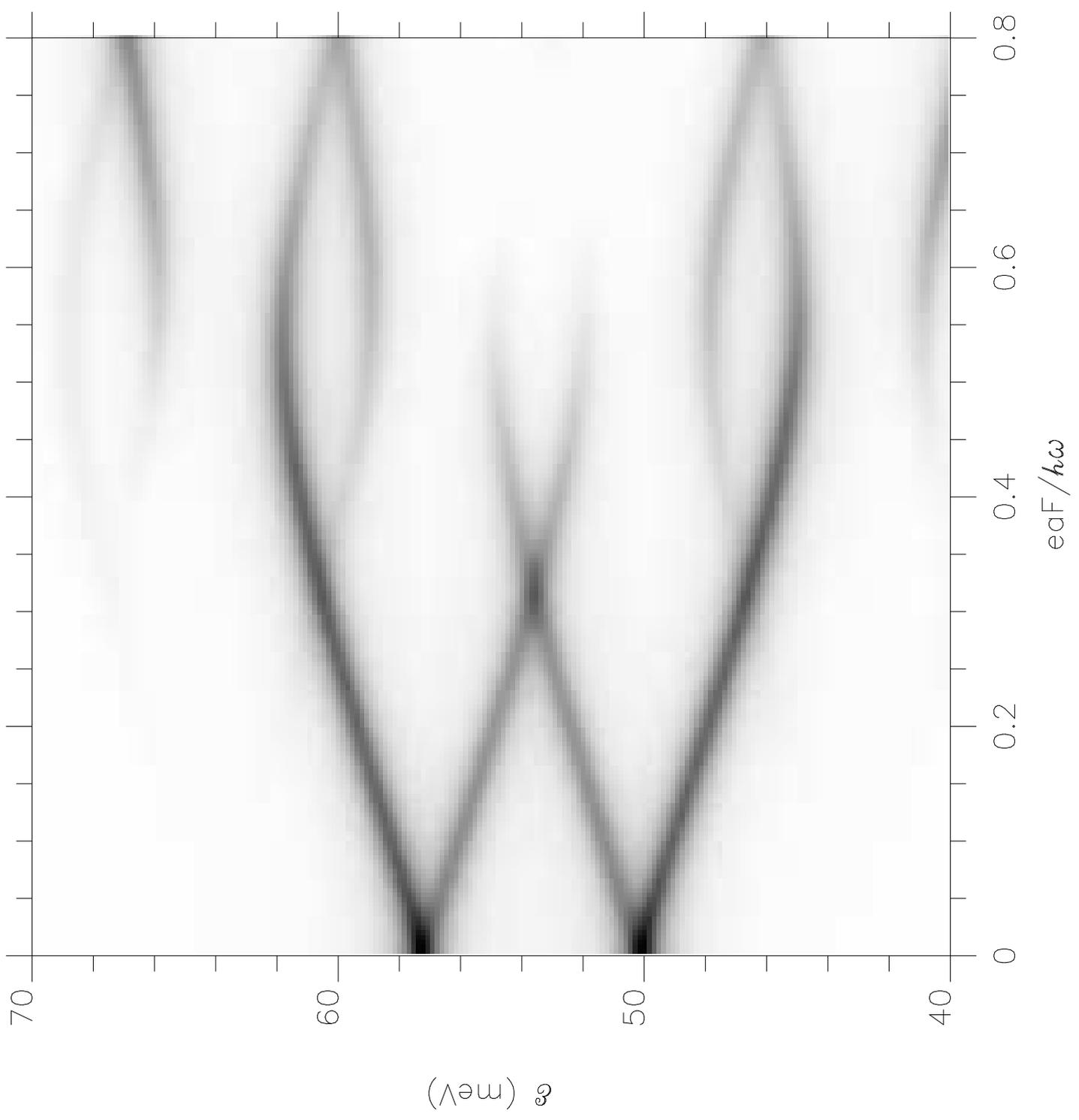
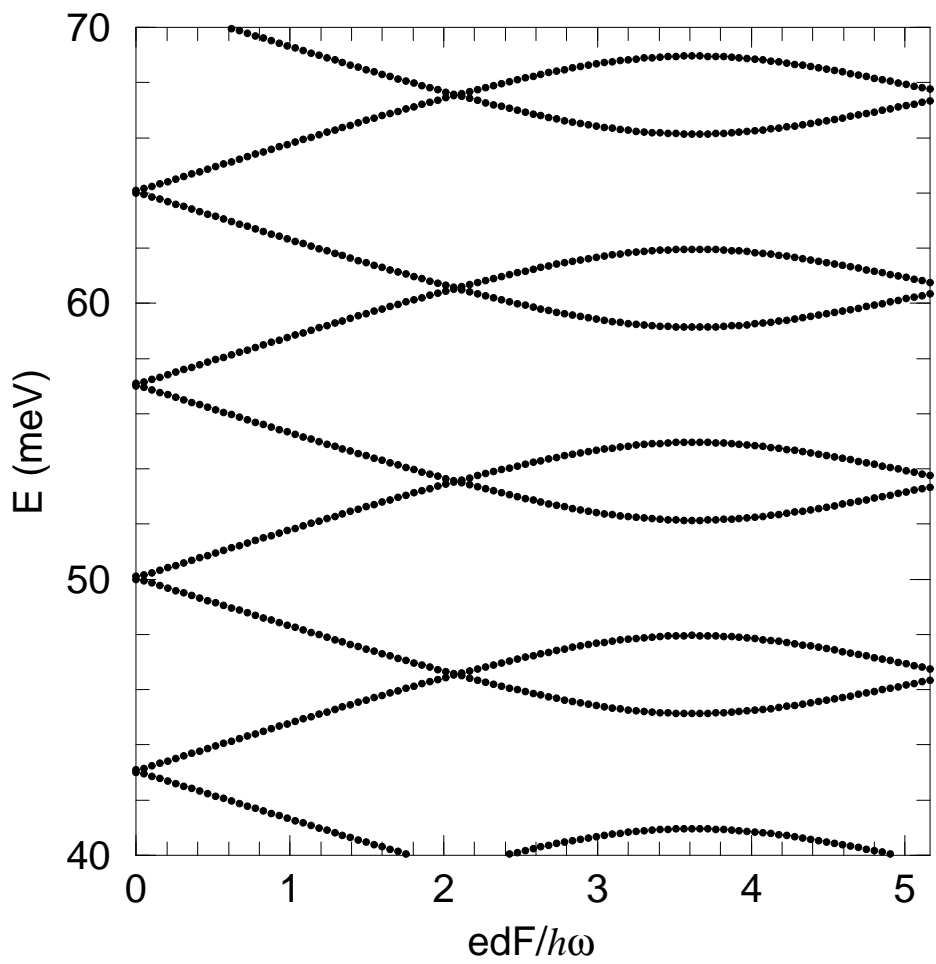
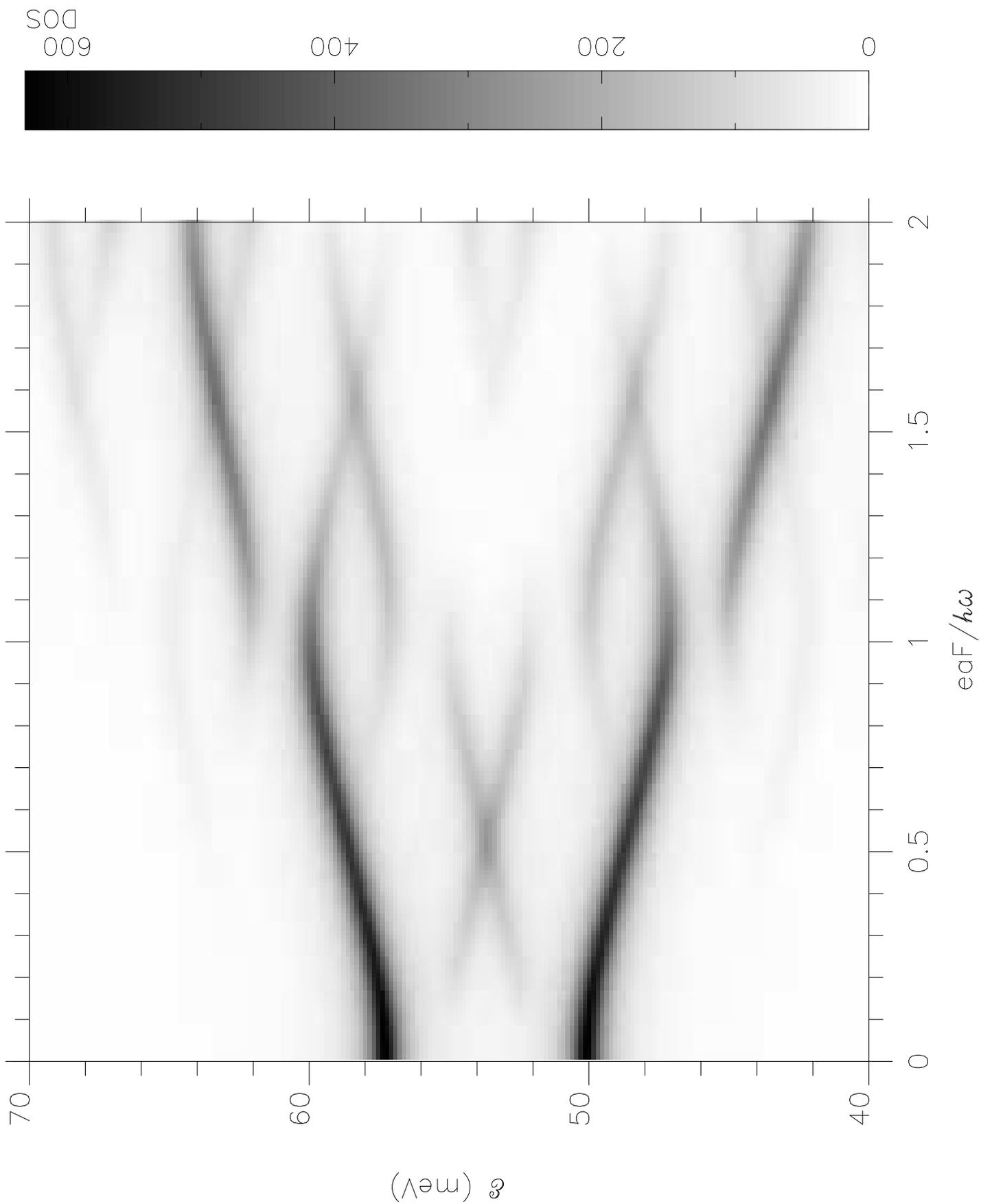
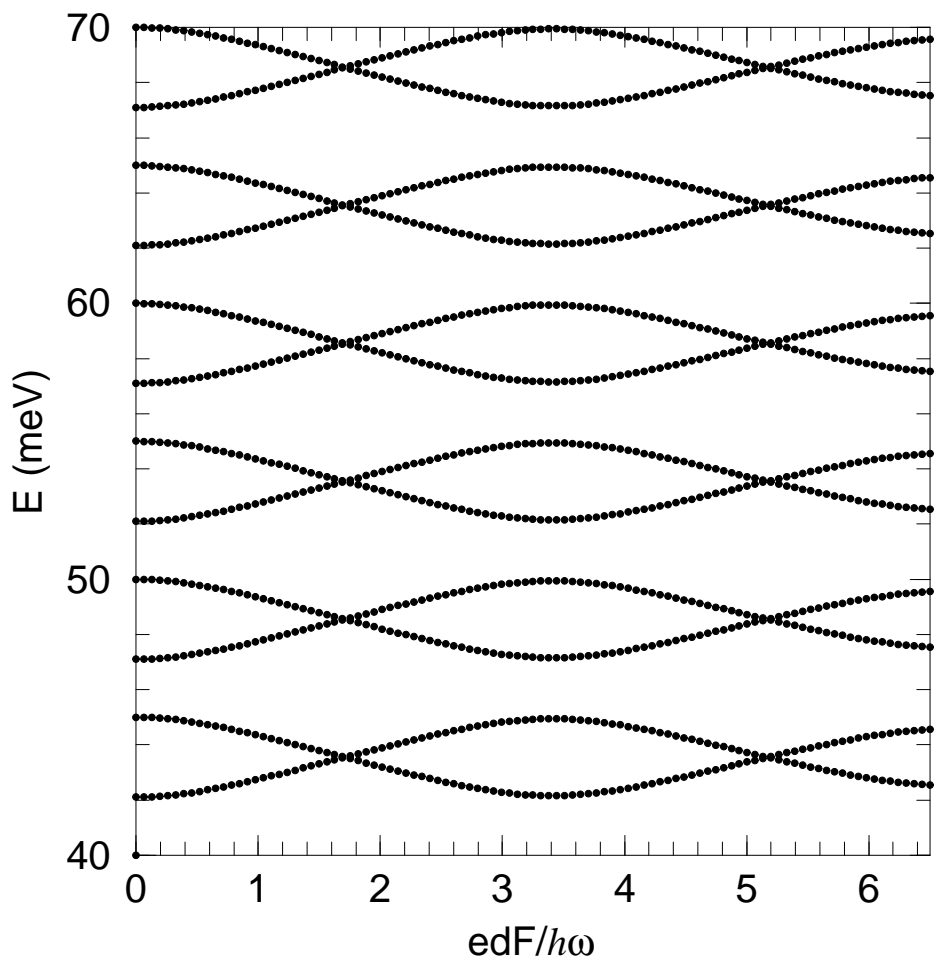
\end{document}